\documentclass[fleqn,10pt]{wlscirep}
\usepackage[utf8]{inputenc}
\usepackage[T1]{fontenc}
\usepackage{lineno}
\linenumbers

\usepackage{longtable}
\newcommand{\xx}{_{\nu xx}}
\newcommand{\xy}{_{\nu xy}}
\newcommand{\xz}{_{\nu xz}}

\newcommand{\yy}{_{\nu yy}}
\newcommand{\yz}{_{\nu yz}}

\newcommand{\zz}{_{\nu zz}}
\newcommand{\icm}{cm$^{-1}$}

\usepackage{mathpazo}

\title{High-throughput computation of Raman spectra from first principles}

\author[1]{Mohammad Bagheri}
\author[1,*]{Hannu-Pekka Komsa}
\affil[1]{Microelectronics Research Unit, Faculty of Information Technology and Electrical Engineering, University of Oulu, Oulu, FIN-90014, Finland}

\affil[*]{corresponding author(s): Hannu-Pekka Komsa (hannu-pekka.komsa@oulu.fi)}

\begin{abstract}
Raman spectroscopy is a widely-used non-destructive material characterization method, which provides information about the vibrational modes of the material and therefore of its atomic structure and chemical composition. Interpretation of the spectra requires comparison to known references and to this end, experimental databases of spectra have been collected. Reference Raman spectra could also be simulated using atomistic first-principles methods but these are computationally demanding and thus the existing databases of computational Raman spectra are fairly small. 
In this work, we developed an optimized workflow to calculate the Raman spectra more efficiently compared to existing approaches. The workflow was benchmarked and validated by comparison to experiments and previous computational methods for select technologically relevant material systems.
Using the workflow, we performed high-throughput calculations for a large set of materials (5099) belonging to many different material classes, and collected the results to a database.
Finally, the contents of database are analyzed and the calculated spectra are shown to agree well with the experimental ones.
\end{abstract}
\begin{document}
\nolinenumbers
\flushbottom
\maketitle

\thispagestyle{empty}

%\noindent Please note: Abbreviations should be introduced at the first mention in the main text – no abbreviations lists or tables should be included. Structure of the main text is provided below.

\section*{Background \& Summary}

%(700 words maximum) An overview of the study design, the assay(s) performed, and the created data, including any background information needed to put this study in the context of previous work and the literature. The section should also briefly outline the broader goals that motivated the creation of this dataset and the potential reuse value. We also encourage authors to include a figure that provides a schematic overview of the study and assay(s) design. The Background \& Summary should not include subheadings. This section and the other main body sections of the manuscript should include citations to the literature as needed.

Raman spectroscopy is a widely used, powerful, and nondestructive tool for analysis and identification of materials as well as assessing material quality. It is based on characterization of the vibrational modes of materials and provides rich atom- or chemical bond-specific information about the crystal structure and chemical composition. When used in assessing material quality, Raman spectra contains information about grain sizes, defect densities, and strain, among others \cite{DAS_2011,schrader2008infrared,parker1983applications,vavskova2011powerful}. In other fields, Raman spectroscopy has been used to, e.g., detect counterfeit medicines, identify plastic types in recycling flows, to detect hazardous chemicals, or to measure temperature \cite{SCOTTER1997285,Bicchieri2006,Orlando2021,Adya2020,Taghizadeh2020}.
Raman spectrum provides a fingerprint of the material, but it is usually not possible to directly interpret e.g. the material composition from the spectrum. In order to use Raman in the above-mentioned material classification and identification applications, a database of known reference spectra is needed. To this end, databases of experimental spectra have been collected, such as the RRUFF Project \cite{Lafuente2016} that contains a large set of experimental Raman spectra of minerals (4112 public samples),
KnowItAll Raman Spectral Library \cite{KnowItAll} that include Raman spectra of different organic and inorganic compounds, polymers and monomers (over 25000 records),
and Raman Open Database(ROD) \cite{El_Mendili_2019} which complements the crystallographic information found in the
Crystallographic Open Database (COD) \cite{cod} (1133 entries).

A Raman Spectrum database made via ab initio, density-functional theory (DFT) electronic structure calculations could be highly useful in providing supplementary information that is difficult to obtain from experiments. For instance, some materials can be difficult to synthesize in a pure form, or their purity or phase content is unknown. The calculated results are also free of any instrumental contributions. Computational studies can also be faster and cheaper to carry out than experiments. Such a database would also be useful to computational researchers, e.g., by providing a reference spectra. Moreover, large datasets can be used in material informatics for material screening or for training models via machine-learning. Still, compared to the experimental ones, the computational databases are of very limited size.
This is due to the computational cost of these calculations, which makes them limited to small systems and/or a small number of materials.
A few open-access libraries of computational Raman spectra already exist such as: (i) Computational 2D Materials Database (C2DB) \cite{Haastrup_2018, Taghizadeh2020} that contains properties of a large number of 2D materials but only 733 structures have Raman spectra, (ii) WURM project \cite{wurm} is a database of computed Raman and infrared spectra for 461 minerals, and (iii) in developing high-throughput computational methods, Liang et al. calculated 55 inorganic compounds \cite{Liang2019}.

In this paper, we report on our research to develop optimized high-throughput workflow to carry out these calculations and build a large database of computational Raman spectra.
For selected systems, the calculated spectra are compared to those obtained using previous computational methods as well as to the experimental ones reported in the literature.
The database of Raman spectra and vibrational properties reported along with this paper consists of 5099 compounds from many different material classes, far surpassing in size the previous computational databases and comparable to the experimental ones.

\section*{Methods}

\subsection*{Simulation of Raman spectra}

In Raman spectroscopy measurements, incident laser photons with a specific frequency $\omega_L$ interact with lattice vibrations, described in the form of phonons in crystalline materials, and the spectrum of inelastically scattered photons are recorded. Scattered photons exhibit either a decrease in frequency $\omega_S$ upon creation of phonon or increase in frequency upon annihilation of a phonon, denoted as Stokes or Anti-Stokes shifts, respectively. 
The intensity of the peaks is related to the Raman scattering cross section, which can be challenging to calculate since the ion (and electron) dynamics in the material need to be described concurrently with the light-matter interaction \cite{Cardona82,Reichardt19_PRB}.

There are several approaches for calculating the Raman spectra: (i) scattering probability from third-order perturbation theory (absorption, electron-phonon coupling, and emission) \cite{Lee_1979,Long2002-qo,Taghizadeh2020}, 
(ii) from the gradient of the electronic susceptibility (usually via finite-differences) in Placzek approximation \cite{placzek1934,Porezag1996,Long2002-qo}, 
and (iii) from the auto-correlation function of time-dependent susceptiblity \cite{GORDON19681,Thomas13}.
Methods (i) and (ii) only yield the Raman tensor, but the phonon eigenvectors and frequencies need to be determined first in a separate calculation step. In method (iii), the peak positions and intensities are obtained at once, but it is computationally highly demanding.
Method (ii) is computationally most affordable and easy to implement in
high-throughput setting \cite{Liang2019} and thus adopted in this work. The method is briefly described below. 

In the first step, the phonons are calculated as described in depth in many previous publications \cite{YuCardona,BaroniRMP}. 
Within harmonic approximation, the potential energy surface is written as a Taylor expansion 
$U=U_0 + \Phi_{\alpha\beta}(ki, lj) u_{\alpha}(ki) u_{\beta}(lj)$, 
where $U_0$ is the ground state energy
and force constant matrix $\Phi$ describes the second-order change in potential energy,

\begin{equation}
\Phi_{\alpha\beta}(ki, lj) = \frac{\partial^2 U}{\partial  u_{\alpha}(ki)\partial  u_{\beta}(lj)} = \frac{\partial F_\alpha(ki)}{\partial u_\beta(lj)}
\label{FC}
\end{equation}
In Eq.~\eqref{FC}, $u_{\alpha}(ki)$ is the displacement of the $k$th atom in the $i$th unit cell in the cartesian direction $\alpha$. $F_\alpha(ki)$ is the force in atom $ki$, and in the equation above its change is induced by the displacement of atom $lj$.
After harmonic ansatz for the temporal evolution of the vibrational modes $v$,
the classical equations of motion for atoms in unit cell "$0$" become

\begin{equation}
M_{k}\omega^2 v_\alpha(k0) = \sum_{l,j,\beta}\Phi_{\alpha,\beta}(k0,lj)v_\beta(lj)
\label{motions}
\end{equation}
where $M_k$ is the mass of atom $k$.
The infinite sums over unit cells $l$ in periodic crystals can be avoided by moving to reciprocal space and, after rescaling $v$ and $\Phi$ by $\sqrt{M}$, Eq.~\ref{motions} is cast into an eigenvalue equation

\begin{equation}
\sum_{l\beta}D_{\alpha \beta}(kl,q)e_{\beta}(l,q\nu) = [\omega(q\nu)]^2 e_{\alpha}(k,q\nu)
\label{eigenvalue}
\end{equation}
where $D$ is the mass-scaled Fourier-transformed $\Phi$ (denoted dynamical matrix), $q$ is the wave vector, $e$ is the eigenvector of the band index $\nu$, and $\omega^2$ are the eigenvalues.
To obtain $D$, force constants $\Phi$ need to be evaluated from the forces induced at atoms $lj$ by displacing each atom $k0$ in the unit cell. To guarantee sufficiently large distance between atoms $k0$ and $lj$, supercell calculations are usually required.
If the crystal symmetry is not considered, the construction of the force constant matrix requires performing $3N$ DFT calculations when each of the $N$ atoms in the unit cell is displaced in each of the three cartesian directions.

Differential cross section for the Stokes component of Raman scattering from the $\nu$th eigenmode far from resonance is given as \cite{Cardona82,Porezag1996}
\begin{equation}
\frac{d\sigma_{\nu}}{d\Omega} = \frac{\omega_S^4 V^2}{(4\pi)^2c^4}\mid \hat{E}_{S}\frac{\partial \chi}{\partial \xi_{\nu}}\hat{E}_{L} \mid^2 \frac{\hbar(n+1)}{2\omega_{\nu}}
\label{Stokes}
\end{equation}
where $\hat{E}_{S}$ and $\hat{E}_{L}$ are the unit vectors of the polarization for the scattered and the incident light, V is scattering volume, and
$\xi$ is a normal-mode coordinate along the mass-scaled eigenvector $e'_\alpha(k) = e_\alpha(k)/\sqrt{M_k} \sim v_\alpha(k)$
and $\chi$ is the electronic susceptibility tensor.
The directional derivative can be written out as
\begin{equation}
\frac{\partial \chi}{\partial \xi}
= \nabla \chi \cdot e'
=\sum_k^{unit cell}\frac{\partial \chi}{\partial u_\alpha(k)} M_k^{-\frac{1}{2}}e_\alpha(k)
\approx \frac{\chi(R_0+ h'e')-\chi(R_0-h'e')}{2h'}
= \frac{\chi(R_0+ h\hat{e'})-\chi(R_0-h\hat{e'})}{2h}|e'|
\label{susceptibility}
\end{equation}
The first two forms involve calculation of derivatives of $\chi$ with respect to displacement of each atom $u(k)$, whereas in the last two forms all atoms are displaced simultaneously along $e'$
and explicitly written in the finite-difference approximation as implemented in the code (displacing the atoms in both positive and negative directions).
Normalized $\hat{e'}=e'/|e'|$ (and $h=h'|e'|$) is used in order to have consistent step size $h$ in systems and modes with different masses (and in units of {\AA}).

Specifically, the Raman tensor is defined as 
\cite{Porezag1996}
\begin{equation}
R_{\nu\beta\gamma} = \frac{V_c}{4\pi} \frac{\partial \chi_{\beta\gamma}}{\partial \xi_\nu}
\label{tensor}
\end{equation}
incorporating $V^2/(4\pi)^2$ from Eq.~\eqref{Stokes}. To evaluate the change in $\chi$, we used the macroscopic dielectric constant $\varepsilon_{\beta\gamma}$ containing only the electronic contribution with clamped ions (sometimes denoted as the high-frequency dielectric constant $\varepsilon_\infty$), which is readily provided by most DFT codes.

While the expression in Eq.~\eqref{Stokes} yields complete information, quite often experimental results are obtained for polycrystalline mineral specimens or powdered samples, in which case the intensity must be averaged over all possible orientations of the crystals.
When the direction of incident light, its polarization, and the direction of outgoing light are all perpendicular, the Raman intensity becomes \cite{Porezag1996,Long2002-qo}
\begin{equation}
\frac{d\sigma_{\nu}}{d\Omega} = \frac{\omega_S^4}{c^4} \frac{\hbar(n+1)}{2\omega_{\nu}} \frac{I_{\rm Raman}}{45}
\label{Stokes2}
\end{equation}
where
\begin{align}
I_{\rm Raman} &= 45 a^2 + 7\gamma^2 \\
a &= \frac{1}{3}(R\xx +R\yy+R\zz) \\
\gamma^2 &= \frac{1}{2}[(R\xx-R\yy)^2 
+(R\xx-R\zz)^2 + (R\yy-R\zz)^2 +6 (R\xy^{2} + R\xz^{2} + R\yz^{2})]
\label{Ramanintensity}
\end{align}
$I_{\rm Raman}$ is Raman activity that is independent of experimental factors such as temperature and incoming photon energy and thus used when comparing our results to other calculations, whereas Eq.\ \ref{Stokes2} is used (and must be used) when comparing to experimental spectra.

\subsection*{Workflow}

We now describe how the theory described above is turned to an efficient computational workflow.
As mentioned, the computational procedure involves two sets of calculations: (i) force constants to get the vibrational modes and (ii) the Raman tensors for each mode. 
While the phonons at $\Gamma$-point can be calculated efficiently, we would like to have access to the full force constant matrix.
This allows calculation of phonon dispersion and also, e.g., estimation of isotope effects and line broadening due to defects or grains via phonon confinement model \cite{Cardona82,Hashemi_2019,Kou2020,Gillet2017}.
Both steps can be computationally demanding for systems with large number of atoms in the unit cell, which has hindered previous efforts to building such databases in the past. 

The most important design decisions that distinguish our work from the previous ones are the following. 
First, we have decided to build our database on top of the Atsushi Togo's Phonon database \cite{Togo2015}, that contains the calculated full force constant matrix, and our work only focuses on calculating the Raman tensors. We are using the same computational parameters, and thus our database is fully consistent with the Phonon database, which is further linked to the Materials project database \cite{mp} via the material-IDs.

Second, to reduce calculation time and make the workflow more efficient compared to existing methods, Raman-active modes are found based on group theory and the Raman tensors are calculated only for modes that are known to be active or whose activity could not be determined. Known inactive modes and the three zero-frequency acoustic modes are ignored. For this purpose, the symmetry information about Raman activity was implemented. 

\begin{figure}[ht]
\centering
\includegraphics[width=15cm]{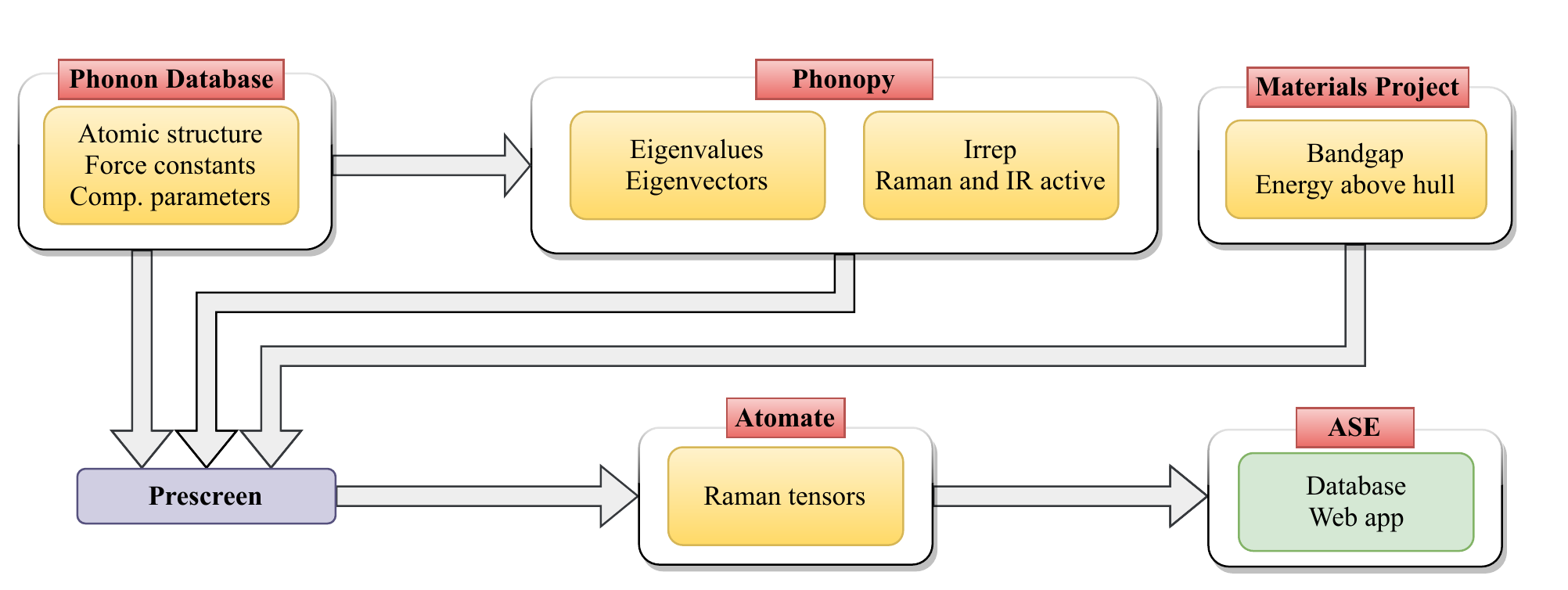}
\caption{High-throughput calculation workflow, grouped according to the software or database used (red box)
and the light yellow boxes indicating the relevant material properties. 
}
\label{fig:wf}
\end{figure}

The workflow developed for automatic Raman tensors calculations is illustrated in Fig.~\ref{fig:wf}.
At the conceptual level, the workflow steps are following:
\begin{enumerate}
    \item Select material from Phonon database, read in optimized structure, computational parameters, and force constant matrix.
    \item Calculate the eigenvectors and eigenvalues at $\Gamma$-point. 
    \item Determine the irreducible representation (irrep) of the modes and whether they are Raman and/or infrared active. 
    \item Perform prescreening to check that the material is dynamically and thermodynamically stable and the material is not metallic or near-metallic.
    \item Calculate the Raman tensors for Raman-active modes and the dielectric tensors for the optimized structure.
    \item All the results (structure, eigenvalues, irreducible representation, Raman tensors, etc.) are collected in a database.
\end{enumerate}

The softwares used in each step are also indicated in Fig.~\ref{fig:wf}.
Atsushi Togo's Phonon database contains the optimized structures, calculated force constants, and all the computational parameters used to obtain them. These are calculated using VASP software \cite{Kresse1996, Kresse_1999}.
The eigenvalues and eigenvectors at $\Gamma$-point, as well as the irreducible representations of the modes are calculated using Phonopy \cite{Togo2015}.
All of this information together with selected material properties obtained from the Materials Project database are collected in a database for prescreening. For this, we adopted to use the database tools in atomic simulation environment (ASE) \cite{Hjorth_Larsen_2017}. In the last step, the calculated Raman tensors are added to this database, which is then also served through a web app implemented in ASE.

For automating the computationally intensive part, i.e., the calculation of the Raman tensors, we used the Atomate \cite{Ceriotti2006} that is a Python-based package for constructing complex materials science computational workflows. The workflow objects generated by Atomate are given to Fireworks workflow software \cite{CPE:CPE3505} for managing, storing, and executing them with the help of Custodian package for error management \cite{ONG2013314}. As the DFT calculator we used here VASP, with the parameters taken from the Phonon database.
During these calculations, all the input parameters and results are stored in a Mongo database, which are afterwards transferred to the database (Computational Raman Database, CRD).

\subsection*{Prescreening}

\begin{figure}[ht]
\centering
\includegraphics[width=7cm]{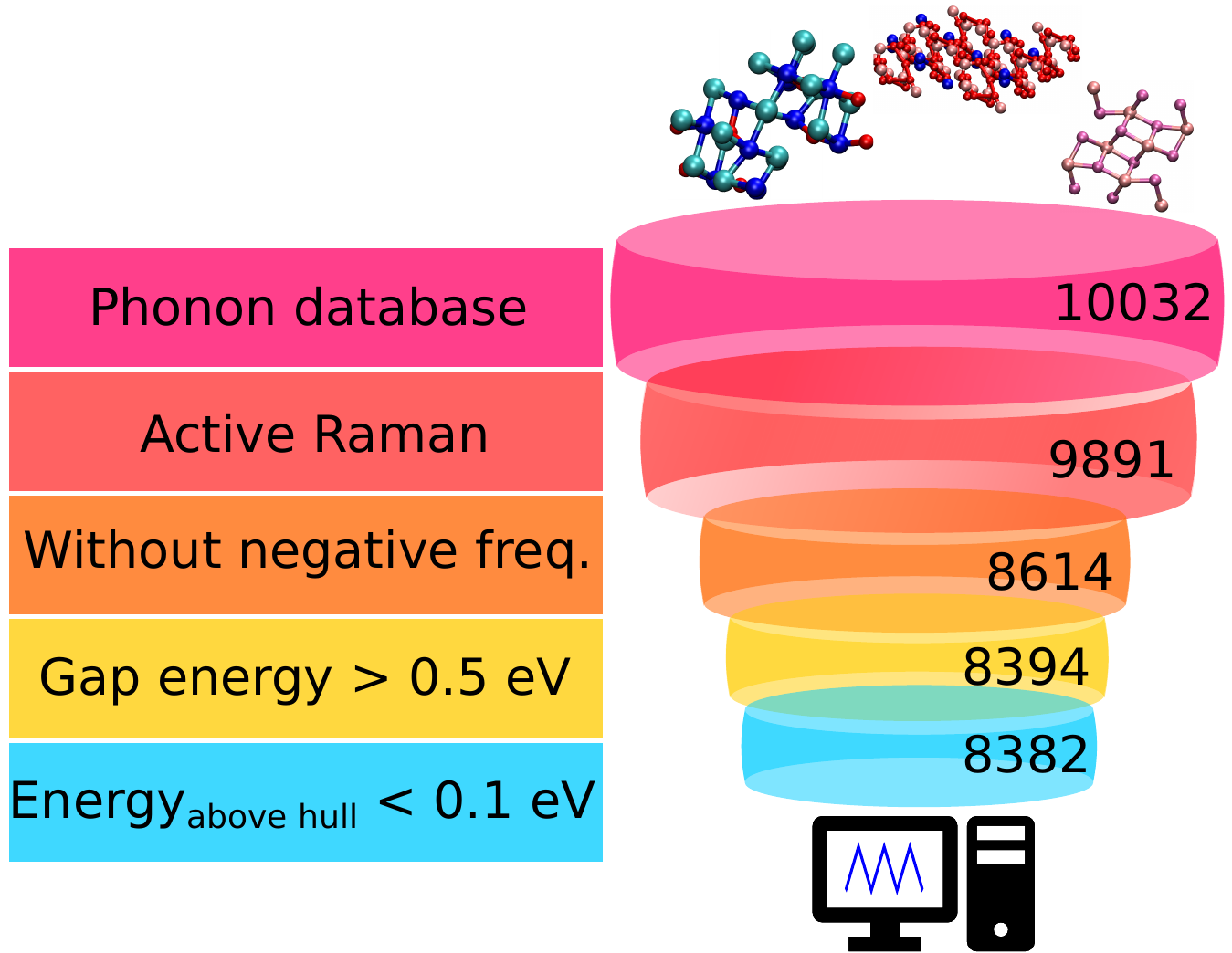}
\caption{Structure selection procedure. The prescreening criteria are indicated on the left and the the number of structures in each step are indicated on the right.}
\label{fig:selected}
\end{figure}

Before Raman tensor calculations we performed the following prescreening, also illustrated in Fig.~\ref{fig:selected}: (i) We check that the material has Raman active mode(s) based on the symmetry analysis.
(ii) We check that the material is dynamically stable, i.e., there are no modes with imaginary frequencies at the $\Gamma$-point. (iii) We check that the material is thermodynamically stable by requiring that the energy above the convex hull is less than 0.1 eV/atom, as materials with the energy $>0.1$ eV are unlikely to be experimentally synthesized\cite{Sun16_SciAdv}. (iv) We check that the bandgap is larger than 0.5 eV, since our computational approach is strictly valid only for non-resonant conditions (i.e., photon energy smaller than the band gap), and metallic systems require very large k-point meshes which will increase the computational cost. For (iii) and (iv) we use information from the Materials Project database at the same material ID \cite{mp}. Finally, we have 8382 (83.55\%) materials satisfying these conditions and flagged for calculation. 
It is also worth noting that Phonon database contains only materials that are non-metallic, non-magnetic, and non-triclinic.

The workflow first performs calculation of dielectric tensors of the optimized structure, which can be compared to that provided in Phonon database.
Additionally, the maximum forces are checked in this step and the calculation terminated if the forces are $>0.001$ eV/{\AA}, but no such case was encountered.

\subsection*{Computational parameters}

All density-functional theory (DFT) calculations are carried out using VASP (Vienna Ab initio Simulation Package) \cite{Kresse1996,PhysRevB.54.11169} with projector-augmented wave method \cite{PhysRevB.50.17953}. PBEsol exchange-correlation functional \cite{PhysRevLett.100.136406} and other computational parameters were taken to be the same as used in Phonon database. In particular, plane wave cutoff is set to 1.3 times the maximum cutoff listed in PAW setups.
In Phonon database, the structures of standardized unit cells are given, whereas we adopt to use the primitive cell in Raman tensor calculations to save computational time. The primitive cell can be readily obtained using Phonopy \cite{Togo2015}.
In the calculation of eigenvectors, non-analytic corrections are not included, as the eigenvectors would then depend also on the direction from which $q \to 0$ is approached and thereby complicate the calculations significantly. Fortunately, this mostly happens for the IR-active modes and less for the Raman-active modes. Moreover, the induced change in eigenvectors and in Raman tensors is expected to be small and the splitting of the modes can be determined a posteriori.

There are then only two parameters left to decide: the k-point mesh and the magnitude of the atomic displacements in evaluation of the Raman tensor by finite differences. 

In Phonon database, the Brillouin zone of the unit cell is sampled by a mesh whose density is defined by the $R_k$ parameter in VASP. We adopt the same approach, but it is worth noting that since we use primitive cell, the exact density and positions of mesh points can be slightly different.
Moreover, metals and small-gap semiconductors usually require higher density k-point mesh than large-gap insulators. 
All calculations in the Phonon database used $R_k=20$, which should be sufficient for the structural optimization of materials included in the database (band gap > 0.5 eV). Determination of Raman tensor may, however, require a higher value.
In order to benchmark this, we selected two materials from the Phonon database with different band gaps:
the largest band gap material among the common III-V semiconductors is AlN (4.05 eV) and Si is a small band gap material (0.85 eV).

As illustrated in Fig. S1, $R_k = 40$ is needed to achieve converged results for dielectric constant and Raman intensity of a small band gap material Si, whereas for a large band gap material AlN $R_k=20$ is sufficient. See Benchmark section in SI for more details.
In our workflow, we have chosen to use the following values: $R_k=20$ for the structures with a band gap more than the 2 eV, $R_k=30$ for band gaps in the range of 1--2 eV, and $R_k=40$ for band gaps smaller than 1 eV.

In order to benchmark the displacement, we chose materials with heavy and light elements, PbO and Cd(HO)$_2$. As shown in Fig.~S2, varying the displacement from 0.001 {\AA} to 0.04 {\AA} (default value being 0.005 {\AA}), we found little change in the Raman tensors or the dielectric constants. Therefore, we chose to use the default value.
Finally, we verified the computational workflow in Atomate by comparing the Raman spectra of few structures to those obtained using VASP\_Raman code \cite{vaspraman}. As shown in Fig.~S3, a good agreement is found. 
We note that Atomate had wrong normalization of eigenvectors which in some cases resulted in overestimation of the Raman intensities, but was fixed in the version used here.

%The Methods should include detailed text describing any steps or procedures used in producing the data, including full descriptions of the experimental design, data acquisition assays, and any computational processing (e.g. normalization, image feature extraction). See the detailed section in our submission guidelines for advice on writing a transparent and reproducible methods section. Related methods should be grouped under corresponding subheadings where possible, and methods should be described in enough detail to allow other researchers to interpret and repeat, if required, the full study. Specific data outputs should be explicitly referenced via data citation (see Data Records and Citing Data, below).Authors should cite previous descriptions of the methods under use, but ideally the method descriptions should be complete enough for others to understand and reproduce the methods and processing steps without referring to associated publications. There is no limit to the length of the Methods section. Subheadings should not be numbered.
%\subsection*{Subsection}
%Example text under a subsection. Bulleted lists may be used where appropriate, e.g.
%\begin{itemize}
%\item First item
%\item Second item
%\end{itemize}

%\subsubsection*{Third-level section}
 
%Topical subheadings are allowed.

\section*{Data Records}

\subsection*{Computational Raman Database}
The final database contains vibrational information and Raman tensors stored in JSON document that can be downloaded directly from the Materials Cloud Archive \cite{mparchive} and queried with a simple python script. The Table~\ref{table:1} shows all the database keys with their related descriptions. 
The data can also be browsed online in Computational Raman Database website (\hyperlink{http://ramandb.oulu.fi/}{ramandb.oulu.fi}). 

\subsection*{Database statistics}

As shown in Fig.~\ref{fig:selected}, there were 10032 materials in the Phonon database and 8382 of them were flagged for calculation. 
Since each structure contains several vibrational modes, the total number of modes in our database was 725163, and 428081 modes of them are Raman active or the activity is unknown. 

\begin{figure}[ht]
\centering
\includegraphics[width=15cm]{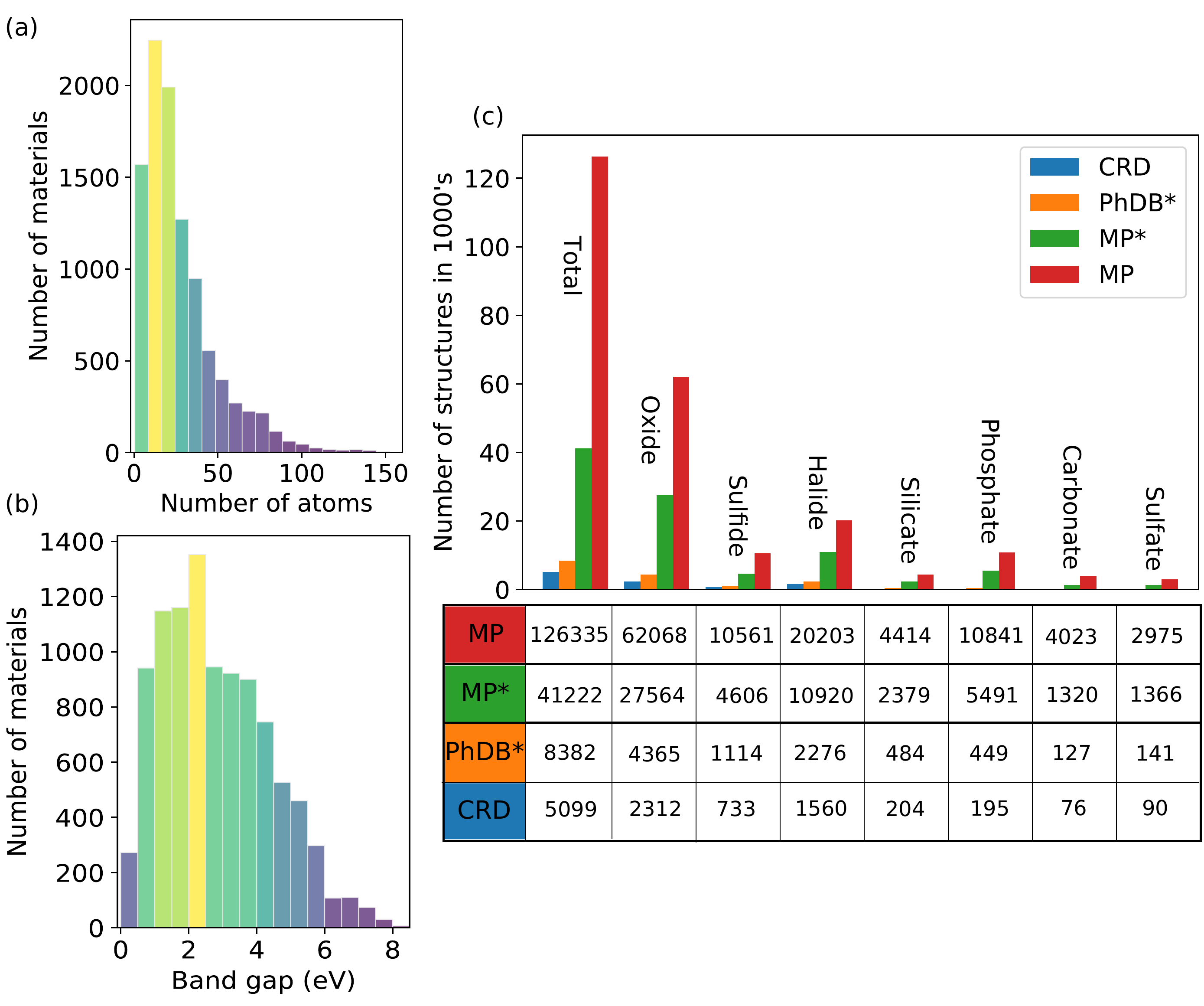}
\caption{Database statistics. (a), (b) The number of materials in Phonon database as a function of number of atoms in structures and band gap, respectively. (c) Comparison of the number of different types of compounds in Material Project (MP) and Computational Raman Database (CRD). MP* and PhDB* shows the number of structures in Materials Project and Phonon database, respectively, when the same selection conditions as in CRD are applied to them.}
\label{fig:mp}
\end{figure}

Figs.~\ref{fig:mp}(a,b) shows the number of materials in the database 
(before prescreening) grouped by the calculated band gaps and the number of atoms in their structures, respectively.
The histogram with respect to the number of atoms, peaks at around 20--30. 
There are some materials with very large primitive cells containing more than 100 atoms, but many of these appear to be disordered/alloyed/defective variants of the small primitive cell systems and thus of limited interest.
Since the Phonon database only includes non-metallic materials, the number of materials with a band gap smaller than 0.5 eV is small, and therefore neglecting those materials in our prescreening step has small impact. 

We proceeded to carry out the Raman tensor calculations in the order of increasing number of atoms in the primitive cell. The database included here contains 5099 calculated structures.
We calculated all materials with less than 10 atoms in the primitive cell and all experimentally observed materials (as indicated by MP) less than 40 atoms in the primitive cell. 
For this, we used about 9.5 million CPU hours. We estimate that for calculating the remaining 3283 structures would require more than 20 million CPU hours, owing to the much larger cell sizes.

In Fig.~\ref{fig:mp}(c) we compare the number of materials considered in this work and in Materials Project database as grouped by the type of compound (oxides, halides, etc.). "MP" denotes the full Materials Project database, whereas "MP*" includes the same conditions (band gap larger than 0.5 eV and energy above hull less than 0.1 eV) as used in our material set (PhDB*). "CRD" refers to the calculated set of materials. In total, almost 20\% of the MP* structures are contained in the PhDB* dataset and about 12\% are calculated.
Also, the different types of compounds are included in our database with similar statistics as in Materials Project. As an example, the percentage of oxides and halogenides are 52 \% and 27 \% in our database, compared to 67 \% and 26 \% in MP*.
Finally, we used the algorithm proposed by Larsen et al. \cite{Larsen_2019} for identifying the dimensionality of the structures in our database: 4137 structures (more than 80 \%) are three-dimensional, 385 structures are two-dimensional, 72 structures are one-dimensional, 277 structures are 0D and others are a mixture of different dimensionality, such as 0D+1D, 0D+2D, 0D+3D, etc.
This shows that our database covers most different material classes.

%The Data Records section should be used to explain each data record associated with this work, including the repository where this information is stored, and to provide an overview of the data files and their formats. Each external data record should be cited numerically in the text of this section, for example \cite{Hao:gidmaps:2014}, and included in the main reference list as described below. A data citation should also be placed in the subsection of the Methods containing the data-collection or analytical procedure(s) used to derive the corresponding record. Providing a direct link to the dataset may also be helpful to readers (\hyperlink{https://doi.org/10.6084/m9.figshare.853801}{https://doi.org/10.6084/m9.figshare.853801}).Tables should be used to support the data records, and should clearly indicate the samples and subjects (study inputs), their provenance, and the experimental manipulations performed on each (please see 'Tables' below). They should also specify the data output resulting from each data-collection or analytical step, should these form part of the archived record.

\section*{Technical Validation}

\subsection*{Comparison to experiments}

Selected computational benchmarks were already presented in the Computational parameters section. 
In this section, we compare the calculated spectra from our approach with experimental results extracted from the RRUFF database to validate our method and calculations. 
RRUFF contains only (estimated) chemical formula and lattice parameters but not atomic positions, and thus we cannot guarantee exact structural match. 
Based on mineral names, there are 703 entries in RRUFF database that matched with 288 structures of our database.
The Table S1 contains mineral names, formula, and their RRUFF IDs for structures with the same formula as found in Phonon database, 92 in total. 
27 of these were found to have the similar lattice parameters compared to the matched structure in our database and thus very likely to be the same structure. Moreover, in most cases, the energy above hull is zero or very small, the maximum being 40 meV/atom.

\begin{figure}[ht]
\centering
\includegraphics[width=15cm]{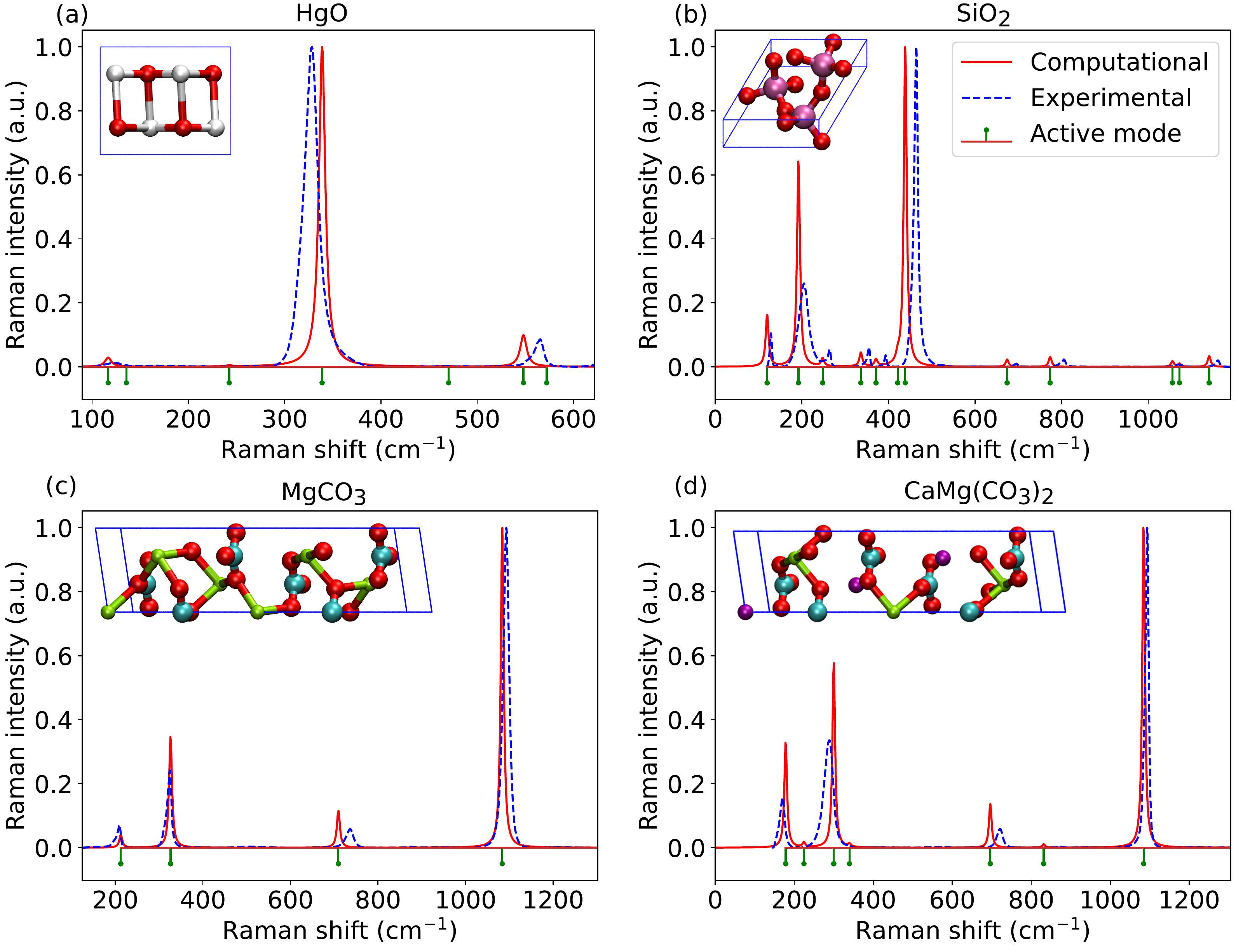}
\caption{Comparison of calculated Raman spectra (red solid line) and experimental spectra from RRUFF database \cite{Lafuente2016} (blue dashed line) for selected minerals. 
Green short line segments show the Raman active modes based on the symmetry analysis. Both spectra are normalized to one at maximum. The experimental spectra correspond to processed data measured at wavelength 532 nm from unoriented samples with RRUFF id: R140877, R050125, R050676, and R050129, for HgO, SiO$_2$, MgCO$_3$, and CaMg(CO$_3$)$_2$, respectively.
The atomic structures are given in the inset (O: red, Hg: white, Si: magenta, Mg: green, C: cyan, Ca: purple).
}
\label{fig:intensity}
\end{figure}

Fig.~\ref{fig:intensity} shows a comparison between calculated spectra and experimental Raman spectra of few selected minerals: HgO, MgCO$_3$, CaMg(CO$_3$)$_2$, and SiO$_2$. 
Overall good agreement between computational and experimental results is found. In most cases, the frequencies (Raman shifts) differ from the experimental values by less than few percent. The variation in peak intensities is somewhat larger but qualitatively correct. We note that the comparison to the experiment is complicated by the varying linewidths in the experimental spectra, which in turn modifies the peak maxima. The linewidth is related to the phonon lifetime, which is not evaluated in our calculations. Instead, in the simulated spectra we have only included a reasonable phonon lifetime-induced broadening of 8 \icm{}. The experimental spectra appear to contain also a Gaussian-type (instrumental) broadening, which we do not attempt to reproduce here.
Also, while perfectly ordered bulk crystals are used in calculations, in experiments the material purity or even exact composition may be unknown and the spectrum is affected by parameters such as temperature, pressure, and measurement geometry.
While we are relying in harmonic approximation, phonon renormalization due to anharmonic effects can affect the frequencies as well as linewidths. Also, we are simulating non-resonant Raman spectra, while in resonant Raman the intensities may change depending on the electronic resonance conditions.
Nevertheless, in the cases where the Raman tensors are affected by any of these effects, the Raman-active modes found based on the group theory can still be used to assist in the analysis of the experimental spectra.

%This section presents any experiments or analyses that are needed to support the technical quality of the dataset. This section may be supported by figures and tables, as needed. This is a required section; authors must present information justifying the reliability of their data.

\section*{Usage Notes}

We introduced an optimized workflow for performing high-throughput first-principles calculations of Raman tensors.
The workflow takes full advantage of the crystal symmetry, adopts carefully benchmarked computational parameters, and avoids calculation of vibrational modes by importing them from existing Phonon database.
We carried out such calculations for 5099 materials and the results are included in the dataset accompanying this paper. 
The database encompasses a wide variety of materials from different compound classes (oxides, halides, etc.) and of different dimensionality. The calculated spectra were also shown to compare favorably with the experimental ones.

The final database contains Raman tensors and other vibrational information, such as phonon eigenmodes, Born charges, and symmetry information, stored in JSON document that can be downloaded directly from the Materials Cloud Archive \cite{mparchive} and queried with a simple python script. 
%The Table~S2 shows all the database keys with their related descriptions. 
The whole dataset can also be browsed online in Computational Raman Database website (\hyperlink{http://ramandb.oulu.fi/}{http://ramandb.oulu.fi}), wherein one can also find other relevant information, such as atomic structure, phonon dispersion, and infrared spectrum.
We hope that the vibrational properties and Raman spectra of materials in the database will prove useful for computational and experimental researchers alike.

%We introduced an optimized workflow for performing high-throughput first-principles calculations of Raman tensors. We carried out such calculations for 5099 compounds and the results are included in the dataset accompanying this paper. 

%The whole dataset can also be browsed online on the Computational Raman database website (\hyperlink{http://ramandb.oulu.fi/}{ramandb.oulu.fi}), which we hope will provide useful information of the vibrational properties and Raman spectra of materials for computational and experimental researchers alike.

%The Usage Notes should contain brief instructions to assist other researchers with reuse of the data. This may include discussion of software packages that are suitable for analysing the assay data files, suggested downstream processing steps (e.g. normalization, etc.), or tips for integrating or comparing the data records with other datasets. Authors are encouraged to provide code, programs or data-processing workflows if they may help others understand or use the data. Please see our code availability policy for advice on supplying custom code alongside Data Descriptor manuscripts.For studies involving privacy or safety controls on public access to the data, this section should describe in detail these controls, including how authors can apply to access the data, what criteria will be used to determine who may access the data, and any limitations on data use. 

\section*{Code availability}

VASP \cite{Kresse1996,Kresse_1999} used in all DFT calculations is a proprietary software. For the database, dimensionality analysis, and web app, we used Atomic Simulation Environment (ASE) \cite{Hjorth_Larsen_2017}, released under GNU Lesser General Public License (LGPL). Phonopy \cite{Togo2015} used in calculating the eigenvectors and performing symmetry analysis is released under New Berkeley Software Distribution (BSD) License. The workflow is defined as a part of Atomate code package \cite{Ceriotti2006} with FireWorks \cite{CPE:CPE3505} for defining, managing, and executing jobs which both are released under a modified BSD license and free to the public. Pymatgen (Python Materials Genomics) used for producing inputs parameters and custodian \cite{ONG2013314} for performing error checking are both open-source packages under Massachusetts Institute of Technology (MIT) license. To store results and task parameters, MongoDB NoSQL database was used with the Server Side Public License (SSPL). 
All the information for prescreening and phonon calculation extracted from Phonon Database \cite{Togo2015,HINUMA2017140} and from Materials project\cite{mp, Ong_2015} are both released under Creative Commons Attribution 4.0 International License.

%For all studies using custom code in the generation or processing of datasets, a statement must be included under the heading "Code availability", indicating whether and how the code can be accessed, including any restrictions to access. This section should also include information on the versions of any software used, if relevant, and any specific variables or parameters used to generate, test, or process the current dataset. 

\bibliography{ref}

%\noindent LaTeX formats citations and references automatically using the bibliography records in your .bib file, which you can edit via the project menu. Use the cite command for an inline citation, e.g. \cite{Kaufman2020, Figueredo:2009dg, Babichev2002, behringer2014manipulating}. For data citations of datasets uploaded to e.g. \emph{figshare}, please use the \verb|howpublished| option in the bib entry to specify the platform and the link, as in the \verb|Hao:gidmaps:2014| example in the sample bibliography file. For journal articles, DOIs should be included for works in press that do not yet have volume or page numbers. For other journal articles, DOIs should be included uniformly for all articles or not at all. We recommend that you encode all DOIs in your bibtex database as full URLs, e.g. https://doi.org/10.1007/s12110-009-9068-2.

\section*{Acknowledgements}

We thank CSC–IT Center for Science Ltd. for generous grants of computer time. Also, We acknowledge discussions with Prof. Atsushi Togo on the details concerning the Phonon database
%Acknowledgements should be brief, and should not include thanks to anonymous referees and editors, or effusive comments. Grant or contribution numbers may be acknowledged.

\section*{Author contributions statement}

%Must include all authors, identified by initials, for example:A.A. conceived the experiment(s), A.A. and B.A. conducted the experiment(s), C.A. and D.A. analysed the results. All authors reviewed the manuscript. 

\section*{Competing interests} %(mandatory statement)

The authors declare no conflict of interest.
%The corresponding author is responsible for providing a \href{https://www.nature.com/sdata/policies/editorial-and-publishing-policies#competing}{competing interests statement} on behalf of all authors of the paper. This statement must be included in the submitted article file.

\section*{Figures \& Tables}

%Figures, tables, and their legends, should be included at the end of the document. Figures and tables can be referenced in \LaTeX{} using the ref command, e.g. Figure \ref{fig:stream} and Table \ref{tab:example}. Authors are encouraged to provide one or more tables that provide basic information on the main ‘inputs’ to the study (e.g. samples, participants, or information sources) and the main data outputs of the study. Tables in the manuscript should generally not be used to present primary data (i.e. measurements). Tables containing primary data should be submitted to an appropriate data repository.Tables may be provided within the \LaTeX{} document or as separate files (tab-delimited text or Excel files). Legends, where needed, should be included here. Generally, a Data Descriptor should have fewer than ten Tables, but more may be allowed when needed. Tables may be of any size, but only Tables which fit onto a single printed page will be included in the PDF version of the article (up to a maximum of three). Due to typesetting constraints, tables that do not fit onto a single A4 page cannot be included in the PDF version of the article and will be made available in the online version only. Any such tables must be labelled in the text as ‘Online-only’ tables and numbered separately from the main table list e.g. ‘Table 1, Table 2, Online-only Table 1’ etc.

\begin{table}[ht]
\centering
\begin{tabular}{|l|l|l|}
 \hline
 Keys & Datatype & Description \\
 \hline
 lattice\_parameters  & list & $a$, $b$, and $c$ lattice constants ({\AA}) \\
 \hline
 lattice\_angles  & list & $\alpha$, $\beta$ and $\gamma$ angles between lattice vectors \\
 \hline
 cell  & array & Lattice vectors in 3$\times$3 matrix format \\
 \hline
 positions & array & Atomic positions in relative coordinates \\
 \hline
 numbers  & array & Total number of atoms and atomic numbers of all elements  \\
  \hline
 mass & float & Sum of atomic masses in the unit cell (amu) \\
 \hline
  volume  & float & Volume of the unit cell ({\AA}$^3$) \\
 \hline
 mpid  & string & MP ID \\
 \hline
 bandgap\_mp & float & Band gap from MP database (eV) \\
 \hline
  bandgap & float & Band gap (eV)\\
 \hline
  cbm & float & Conduction band minimum (eV) \\
 \hline
  vbm & float & Valence band maximum (eV) \\
 \hline
 diel\_mp  & array & Dielectric tensor (electronic contribution) from MP database  \\
 \hline
  diel  & array & Dielectric tensor (electronic contribution) \\
 \hline
 frequencies\_thz  & list & $\Gamma$-point frequencies (THz)  \\
 \hline
  frequencies\_cm  & list & $\Gamma$-point frequencies (1/cm)  \\
 \hline
 pointgroup  & string & Point group \\
 \hline
 spacegroup  & string & Space group \\
 \hline
 chemical\_formula  & string & Chemical formula \\
 \hline
 IRactive  & array & Infrared-active modes  \\
 \hline
 IRlabels  & list & Irreducible representation (irrep) labels of modes  \\
 \hline
 IRbands  & list & Irrep. band groups of degenerate modes \\
 \hline
 natom  & integer & Total number of atoms \\
 \hline
 Ramanactive  & array & Raman activity of modes (0: inactive, 1: active, -1:unknown) \\
 \hline
 raman\_tensors  & array & Raman tensors \\
 \hline
 born  & array & Born charges ($e$) \\
 \hline
 eigenvec & array & Eigenvectors \\
 \hline
 dimensionality  & string & Dimensionality of structure \\
 \hline
 mp\_e\_above\_hull  & float & Energy above convex hull from MP database (eV/atom) \\
 \hline
 negative\_freq\_Gamma  & boolean & Existence of negative frequencies at $\Gamma$-point  \\
 \hline
 negative\_freq\_path  & boolean & Existence of negative frequencies in phonon dispersion \\
 \hline 
  Refs  & string & Links to Phonon database and MP websites  \\
 \hline
\end{tabular}
\caption{\label{table:1}Description of the JSON file structure for Computational Raman Database (CRD)}
\end{table}

\end{document}

\nolinenumbers
\flushbottom
\maketitle

\section*{Benchmark}

To benchmark of our approach, we selected four materials with different band gaps and different atomic masses: Si, PbO, AlN, and Cd(HO)$_2$.
We first verified that the standard approach of calculating Raman-tensors for all modes agrees with the Raman-active modes identified using group theory. That is, all the Raman-inactive modes were found to have vanishingly small Raman tensor, although often nonzero due to numerical errors.

\begin{figure}[ht]
    \centering
    \includegraphics[width=15cm]{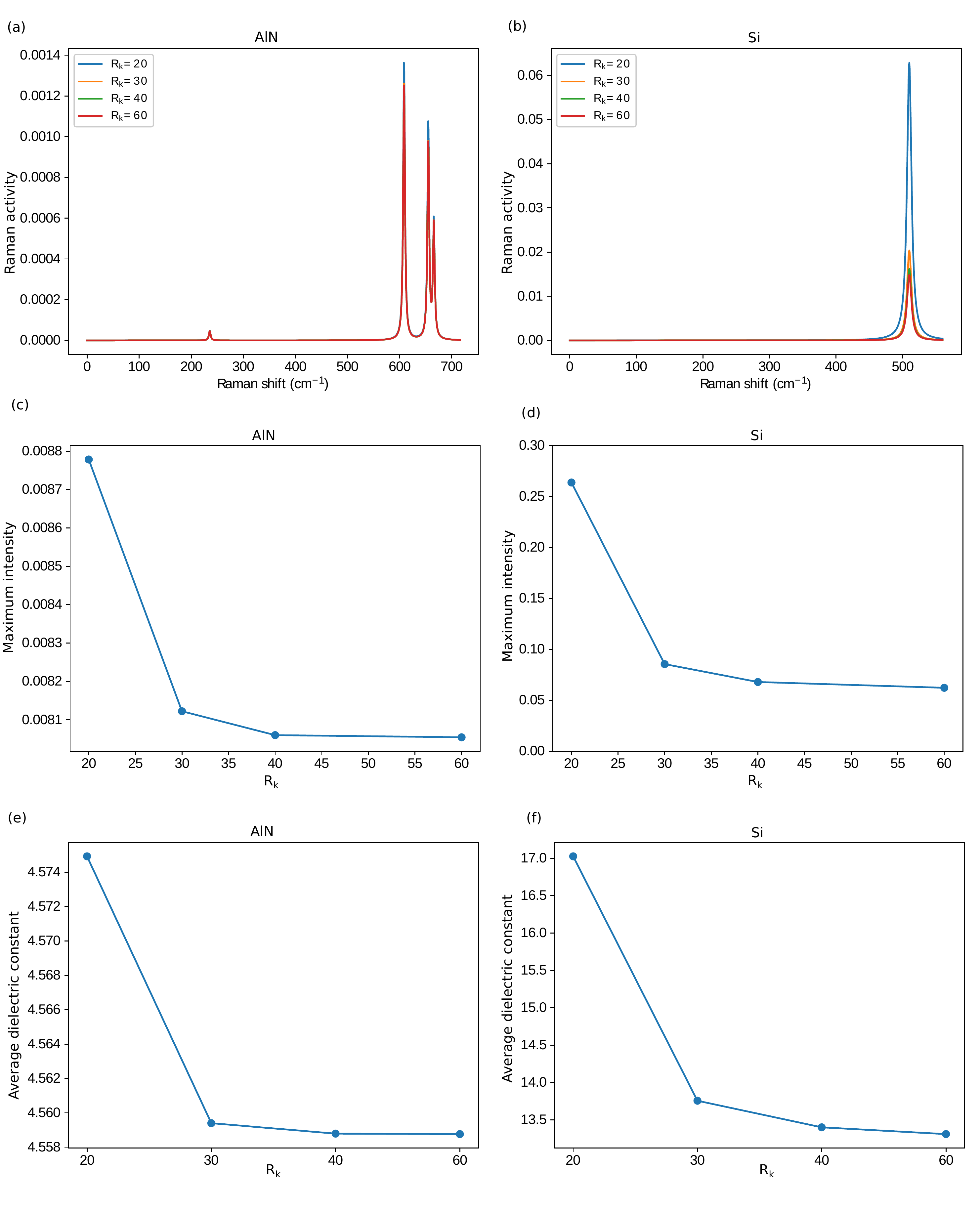}
    \caption{The effect of k-point mesh parameter R$_k$. (a) and (b) show the unnormalized Raman activity of AlN and Si, respectively, with R$_k$ = 20 (blue line), 30 (orange line), 40 (green line) and 60 (red line). (c) and (d) show the effects of different R$_k$ on the maximum intensity of AlN and Si spectra, respectively. (e) and (f) show the effects of different R$_k$ on the average dielectric constant of AlN and Si, respectively.}
    \label{fig:convergance}
\end{figure}

Next, we investigated the effect of k-point mesh density on Raman tensors and Raman spectra. We calculated Raman tensors for R$_k$ = 30, 40 and 60 and compared the results with the standard value (R$_k$ = 20) used in Phonon database. Results from two (out of four) materials are shown in Fig. \ref{fig:convergance}, where, to represent large and small bandgap materials, we selected AlN (4.05 eV) and Si (0.85 eV), respectively. 
Fig. \ref{fig:convergance}(a,b) shows the unnormalized Raman activity spectra for different R$_k$, which clearly illustrates that AlN is hardly affected whereas Si experiences significant changes, thus suggesting that R$_k$ should be increased. To better illustrate the magnitude of changes,
Fig. \ref{fig:convergance}(c,d) show how the maximum intensity changes with increasing R$_k$ and Fig. \ref{fig:convergance}(e,f) shows the average dielectric constant.
In the case of AlN, R$_k$ = 20 already yields Raman tensors within 10 \% of the converged value and dielectric constant within 1 \%.
In the case of Si, R$_k$ = 40 is required to reach similar accuracy.
Based on these results we decided to use R$_k$=40 for materials with bandgap smaller than 1 eV, R$_k$=30 for materials with bandgap between 1 eV to 2 eV, and R$_k$=20 for materials with a bandgap greater than 2 eV.

\begin{figure}[ht]
    \centering
    \includegraphics[width=13cm]{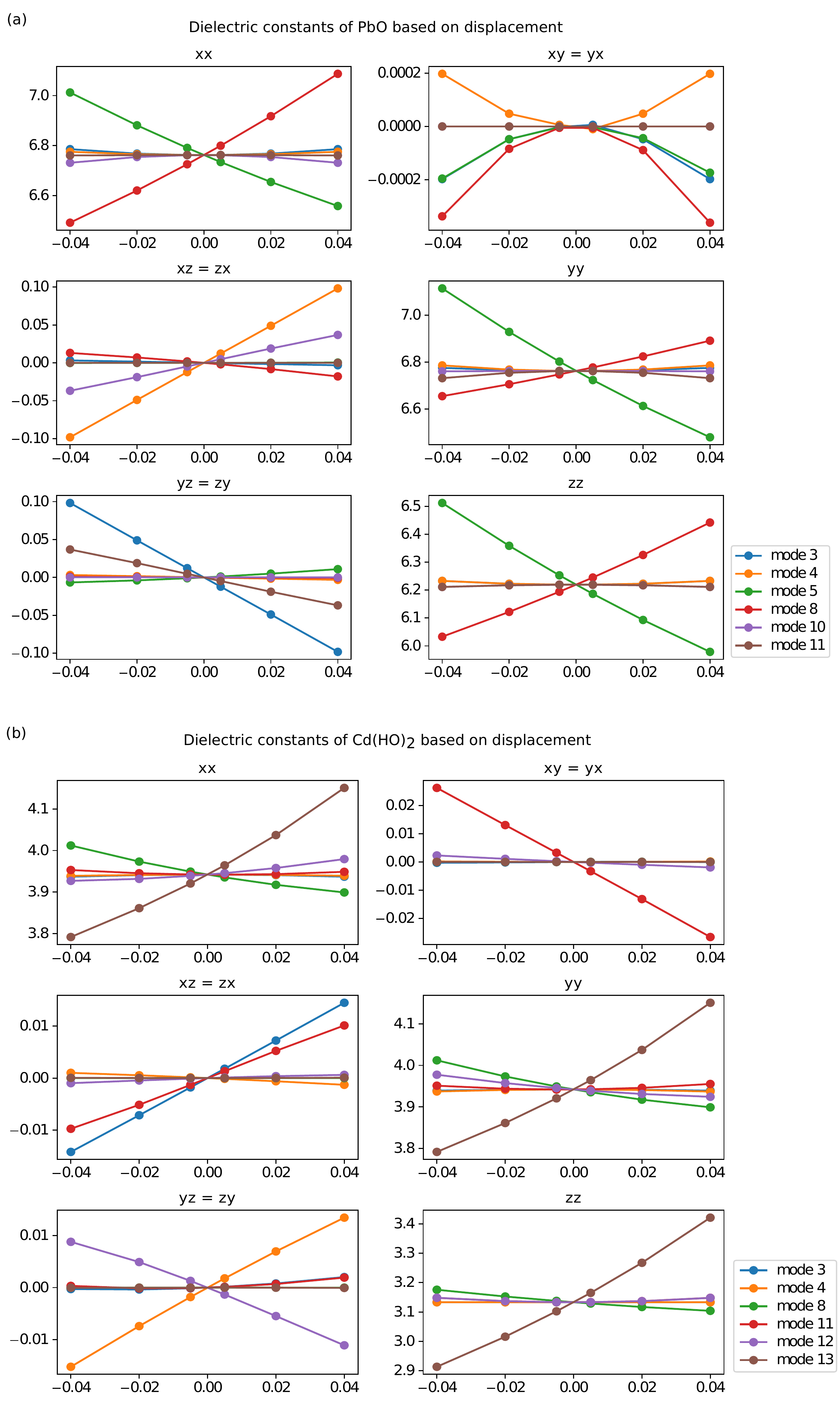}
    \caption{Changes of dielectric constant of (a) PbO and (b) Cd(HO)$_2$ in different directions as a function of the displacement step size (0.005--0.04). }
    \label{fig:dielectric}
\end{figure}

In the third step, we investigated the effect of step size by calculating Raman tensors of PbO and Cd(HO)$_2$ with step sizes of 0.001, 0.02, and 0.04 {\AA} and compared them with those using the standard step size of 0.005 {\AA}. Fig. \ref{fig:dielectric} shows the changes in dielectric constants of PbO and Cd(HO)$_2$ in different directions and plotted for three-step sizes: 0.005, 0.02, and 0.04 {\AA}. Since dielectric tensor is symmetric (xy=yx, xz=zx, and yz=zy), we only plot the inequivalent components. As shown in Fig. \ref{fig:dielectric}, whenever there are pronounced changes in the dielectric constant (corresponding to non-zero components in Raman tensor), the dependence on step size is close to linear. In some cases there is a small parabolic dependence, seen particularly well in the xy=yx component which contains no linear dependence, but these will not affect the Raman tensor since we are using two-point finite-difference stencil. Moreover, in this range of step sizes there is no discernible noise, although some noise could be observed in 0.001 {\AA} results (not shown). Thus, we consider the default value of 0.005 {\AA} a good choice.

\begin{figure}[ht]
\centering
\includegraphics[width=15cm]{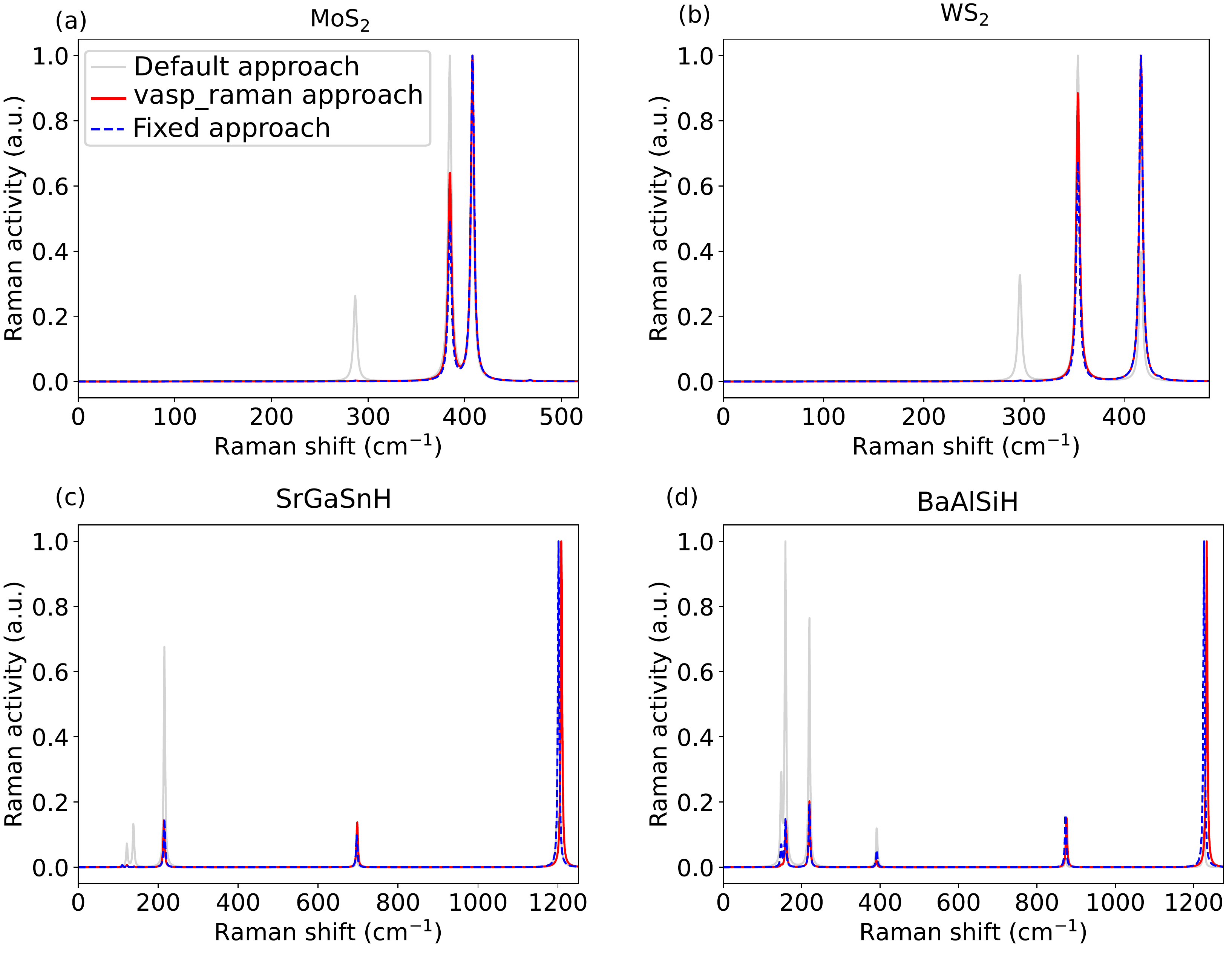}
\caption{Comparison of the Raman activity from our workflow and that from vasp\_raman code. The spectra from old version of Atomate with incorrect eigenvector normalization is also shown.}
\label{fig:verify}
\end{figure}

As mentioned in the text, there was an error in the normalization of eigenvectors in Atomate. We fixed the normalization error and changed the formulations to match with the vasp\_raman code \cite{vaspraman,Porezag1996}. To verify our approach, we used vasp\_raman code to calculate Raman tensors and compared them to Atomate with the fixed and old versions of eigenvector normalization. Fig. \ref{fig:verify} shows the Raman activity spectra of MoS$_2$, WS$_2$, SrGaSnH, and BaAlSiH. The revised normalization yields activities closely matching with vasp\_raman code. The incorrect normalization, on the other hand, tends to lead to overestimation of Raman activities and is particularly severe with modes that have very small Raman activity.

In CRD website (\hyperlink{http://ramandb.oulu.fi/}{ramandb.oulu.fi}), the total Raman intensity is separated into depolarized ($I_\perp$) and polarized ($I_{||}$) components, $I$ = $I_\perp$ + $I_{||}$, with
\begin{align}
\frac{I_{||}}{(\omega_L - \omega_\nu)^4} &\sim \frac{\hbar(n+1)}{30\omega_\nu}(10G_{\nu}^{(0)} +4G_{\nu}^{(2)})
\label{polarized} \\
\frac{I_\perp}{(\omega_L - \omega_\nu)^4} &\sim \frac{\hbar(n+1)}{30\omega_\nu}(5G_{\nu}^{(1)} +3G_{\nu}^{(2)})
\label{depolarized}
\end{align}
where we have taken out the $(\omega_L - \omega_\nu)^4$ term that depends on the laser wavelength, since
(i) this removes the dependence on one external parameter from our spectra,
(ii) our calculations are for non-resonant conditions and one needs to be careful to only compare to wavelengths that are far from resonance,
and (iii) the dependence on $\omega_\nu$ and thus the changes in the spectra after normalization are usually small.
The rotation invariants are \cite{Prosandeev2005,Long2002-qo}
\begin{align}
G_{\nu}^{(0)} &= \frac{1}{3}(R\xx + R\yy + R\zz)^2 
\label{Gi0} \\
G_{\nu}^{(1)} &= \frac{1}{2}[(R\xy-R\yx)^2+(R\xz-R\zx)^2+(R\zy-R\yz)^2]
\label{Gi1} \\
G_{\nu}^{(2)} &= \frac{1}{2}[(R\xy+R\yx)^2+(R\xz+R\zx)^2+(R\zy+R\yz)^2] \nonumber \\ 
&+ \frac{1}{3}[(R\xx-R\yy)^2+(R\xx-R\zz)^2+(R\zz-R\yy)^2]
\label{Gi2}
\end{align}

\begin{comment}
\begin{table}[ht]
\centering
\begin{tabular}{|l|l|l|}
 \hline
 Keys & Datatype & Description \\
 \hline
 cell  & array & lattice parameters  \\
 \hline
 positions & array & Atomic positions  \\
 \hline
 numbers  & arrays & The number of atoms and chemical elements atomic numbers  \\
 \hline
 mpid  & string & Materials Project id  \\
 \hline
 bandgap\_mp & float & Structure's bandgap based on the Materials Project database \\
 \hline
 diel\_mp  & array & Dielectrics based on the Material Project database  \\
 \hline
 frequencies  & array & Gamma point frequencies in THz  \\
 \hline
 pointgroup  & string & Point group of structure  \\
 \hline
 IRactive  & array & IR active modes  \\
 \hline
 IRbands  & array & IR bands  \\
 \hline
 IRlabels  & array & IR labels of modes  \\
 \hline
 Ramanactive  & array & Information about modes that are active or not  \\
 \hline
\end{tabular}
\caption{\label{table:1}Phonon Database Keys and description}
\end{table}
\end{comment}

\begin{longtable}{|l|l|l|l|l|}

\hline \multicolumn{1}{|c|}{Mineral name} & \multicolumn{1}{c|}{Formula} & \multicolumn{1}{c|}{mpid} & \multicolumn{1}{c|}{Energy above hull (eV)} & \multicolumn{1}{c|}{RRUFF ID}\\  \hline 
\endfirsthead

\multicolumn{3}{c}%
{{\bfseries \tablename\ \thetable{} -- continued from previous page}} \\
\hline \multicolumn{1}{|c|}{Mineral name} & \multicolumn{1}{c|}{Formula} & \multicolumn{1}{c|}{mpid} & \multicolumn{1}{c|}{Energy above hull (eV)} & \multicolumn{1}{c|}{RRUFF ID}\\ \hline 
\endhead

\hline \multicolumn{5}{|r|}{{Continued on next page}} \\ \hline
\endfoot

\endlastfoot

Billingsleyite	 & 	Ag$_7$AsS$_6$	 & 	mp-15077	 & 	0.003	 & 	\textbf{R070350}	 \\
\hline
Sanbornite	 & 	BaSi$_2$O$_5$	 & 	mp-3031	 & 	0	 & 	\textbf{R060489}	 \\
\hline
Hardystonite	 & 	Ca$_2$ZnSi$_2$O$_7$	 & 	mp-6227	 & 	0.015	 & 	\textbf{R040026}	 \\
\hline
Perovskite	 & 	CaTiO$_3$	 & 	mp-4019	 & 	0	 & 	\textbf{R050456}	 \\
\hline
Greenockite	 & 	CdS	 & 	mp-672	 & 	0	 & 	\textbf{R090045}	 \\
\hline
Cobaltite	 & 	CoAsS	 & 	mp-4627	 & 	0.001	 & 	\textbf{R070372}	 \\
\hline
Cobaltite	 & 	CoAsS	 & 	mp-16363	 & 	0.004	 & 	\textbf{R060907}	 \\
\hline
Cuprite	 & 	Cu$_2$O	 & 	mp-361	 & 	0	 & 	\textbf{R050374}	 \\
\hline
Stromeyerite	 & 	CuAgS	 & 	mp-5014	 & 	0.024	 & 	\textbf{R060908}	 \\
\hline
Emplectite	 & 	CuBiS$_2$	 & 	mp-22982	 & 	0	 & 	\textbf{R070307}	 \\
\hline
Chalcostibite	 & 	CuSbS$_2$	 & 	mp-4468	 & 	0	 & 	\textbf{R060262}	 \\
\hline
Pyrite	 & 	FeS$_2$	 & 	mp-226	 & 	0.008	 & 	\textbf{R050070}	 \\
\hline
Marcasite	 & 	FeS$_2$	 & 	mp-1522	 & 	0	 & 	\textbf{R060882}	 \\
\hline
Langbeinite	 & 	K$_2$Mg$_2$(SO$_4$)$_3$	 & 	mp-6299	 & 	0	 & 	\textbf{R070285}	 \\
\hline
Aphthitalite	 & 	K$_3$Na(SO$_4$)$_2$	 & 	mp-22457	 & 	0	 & 	\textbf{R050651}	 \\
\hline
Goldschmidtite	 & 	KNbO$_3$	 & 	mp-7375	 & 	0	 & 	\textbf{R190009}	 \\
\hline
Nordite-(La)	 & 	Na$_3$SrLaZnSi$_6$O$_{17}$	 & 	mp-13726	 & 	0	 & 	\textbf{R140310}	 \\
\hline
Swedenborgite	 & 	NaBe$_4$SbO$_7$	 & 	mp-8075	 & 	0	 & 	\textbf{R060486}	 \\
\hline
Leucophanite	 & 	NaCaBeSi$_2$O$_6$F	 & 	mp-560721	 & 	0	 & 	\textbf{R050004}	 \\
\hline
Neighborite	 & 	NaMgF$_3$	 & 	mp-2955	 & 	0	 & 	\textbf{R080108}	 \\
\hline
Cotunnite	 & 	PbCl$_2$	 & 	mp-23291	 & 	0.006	 & 	\textbf{R060655}	 \\
\hline
Matlockite	 & 	PbClF	 & 	mp-22964	 & 	0	 & 	\textbf{R140538}	 \\
\hline
Laurite	 & 	RuS$_2$	 & 	mp-2030	 & 	0	 & 	\textbf{R110120}	 \\
\hline
Zincite	 & 	ZnO	 & 	mp-2133	 & 	0	 & 	\textbf{R060027}	 \\
\hline
Chrysoberyl	 & 	BeAl$_2$O$_4$	 & 	mp-3081	 & 	0	 & 	\textbf{R040073}	 \\
\hline
Wurtzite	 & 	ZnS	 & 	mp-10281	 & 	0.002	 & 	\textbf{R130069}	 \\
\hline
Montroydite	 & 	HgO	 & 	mp-1224	 & 	0	 & 	\textbf{R070235}	 \\
\hline
Quartz	 & 	SiO$_2$	 & 	mp-7000	 & 	0.011	 & 	\textbf{R050125}	 \\
\hline
Bromellite	 & 	BeO	 & 	mp-2542	 & 	0	 & 	X050194	 \\
\hline
Litharge	 & 	PbO	 & 	mp-19921	 & 	0.001	 & 	R060959	 \\
\hline
Romarchite	 & 	SnO	 & 	mp-2097	 & 	0	 & 	R080006	 \\
\hline
Anatase	 & 	TiO$_2$	 & 	mp-390	 & 	0.006	 & 	R060277	 \\
\hline
Andalusite	 & 	Al$_2$SiO$_5$	 & 	mp-4753	 & 	0	 & 	R050258	 \\
\hline
Anglesite	 & 	Pb(SO$_4$)	 & 	mp-3472	 & 	0	 & 	R040004	 \\
\hline
Aragonite	 & 	CaCO$_3$	 & 	mp-4626	 & 	0.024	 & 	R040078	 \\
\hline
Baryte	 & 	Ba(SO$_4$)	 & 	mp-3164	 & 	0	 & 	R040036	 \\
\hline
Brenkite	 & 	Ca$_2$CO$_3$F$_2$	 & 	mp-6246	 & 	0.028	 & 	R060247	 \\
\hline
Calcite	 & 	CaCO$_3$	 & 	mp-3953	 & 	0	 & 	R040070	 \\
\hline
Cerussite	 & 	Pb(CO$_3$)	 & 	mp-19893	 & 	0	 & 	R040069	 \\
\hline
Colquiriite	 & 	CaLiAlF$_6$	 & 	mp-1224	 & 	0	 & 	R070417	 \\
\hline
Dolomite	 & 	CaMg(CO$_3$)$_2$	 & 	mp-6459	 & 	0	 & 	R050129	 \\
\hline
Eitelite	 & 	Na2Mg(CO$_3$)$_2$	 & 	mp-6026	 & 	0	 & 	R110214	 \\
\hline
Eulytine	 & 	Bi$_4$(SiO$_4$)$_3$	 & 	mp-23331	 & 	0	 & 	R060058	 \\
\hline
Farringtonite	 & 	Mg$_3$(PO$_4$)$_2$	 & 	mp-14396	 & 	0	 & 	R130127	 \\
\hline
Geikielite	 & 	MgTiO$_3$	 & 	mp-3771	 & 	0	 & 	R070479	 \\
\hline
Glauberite	 & 	Na$_2$Ca(SO$_4$)$_2$	 & 	mp-6397	 & 	0	 & 	R050350	 \\
\hline
Huntite	 & 	CaMg$_3$(CO$_3$)$_4$	 & 	mp-6524	 & 	0.004	 & 	R040126	 \\
\hline
Cristobalite	 & 	SiO$_2$	 & 	mp-6945	 & 	0.003	 & 	R070235	 \\
\hline
Leiteite	 & 	ZnAs$_2$O$_4$	 & 	mp-29509	 & 	0.006	 & 	R040011	 \\
\hline
Lithiophosphate	 & 	Li$_3$(PO$_4$)	 & 	mp-2878	 & 	0.001	 & 	R100092	 \\
\hline
Magnesite	 & 	Mg(CO$_3$)	 & 	mp-5348	 & 	0	 & 	R040114	 \\
\hline
Nahcolite	 & 	NaH(CO$_3$)	 & 	mp-696396	 & 	0	 & 	R070237	 \\
\hline
Witherite	 & 	Ba(CO$_3$) & 	mp-5504	 & 	0	 & 	R040040	 \\
\hline
Arsenolite	 & 	As$_2$O$_3$	 & 	mp-2184	 & 	0.009	 & 	R050383	 \\
\hline
Åkermanite	 & 	Ca$_2$MgSi$_2$O$_7$	 & 	mp-6094	 & 	0.023	 & 	R061085	 \\
\hline
Benitoite	 & 	BaTi(SiO$_3$)$_3$	 & 	mp-6661	 & 	0	 & 	R050320	 \\
\hline
Gahnite	 & 	ZnAl$_2$O$_4$	 & 	mp-2908	 & 	0	 & 	R070591	 \\
\hline
Rosiaite	 & 	PbSb$_2$O$_6$	 & 	mp-20727	 & 	0	 & 	R070384	 \\
\hline
Xanthoconite	 & 	Ag$_3$AsS$_3$	 & 	mp-561620	 & 	0	 & 	R070746	 \\
\hline
Topaz	 & 	Al$_2$SiO$_4$F$_2$	 & 	mp-6280	 & 	0	 & 	R040121	 \\
\hline
Imiterite	 & 	Ag$_2$HgS$_2$	 & 	mp-9635	 & 	0.03	 & 	R080014	 \\
\hline
Acanthite	 & 	Ag$_2$S	 & 	mp-610517	 & 	0.024	 & 	R070578	 \\
\hline
Argyrodite	 & 	Ag$_8$GeS$_6$	 & 	mp-9770	 & 	0	 & 	R050437	 \\
\hline
Andalusite	 & 	Al$_2$SiO$_5$	 & 	mp-4934	 & 	0.007	 & 	R050258	 \\
\hline
Nitrobarite	 & 	Ba(NO$_3$)$_2$	 & 	mp-4396	 & 	0	 & 	R060622	 \\
\hline
Barylite	 & 	BaBe$_2$Si$_2$O$_7$	 & 	mp-6383	 & 	0	 & 	R060620	 \\
\hline
Barylite	 & 	BaBe$_2$Si$_2$O$_7$	 & 	mp-12797	 & 	0	 & 	R060606	 \\
\hline
Guanajuatite	 & 	Bi$_2$Se$_3$	 & 	mp-23164	 & 	0.028	 & 	R080140	 \\
\hline
Merwinite	 & 	Ca$_3$Mg(SiO$_4$)$_2$	 & 	mp-558209	 & 	0.038	 & 	R070195	 \\
\hline
Rankinite	 & 	Ca$_3$Si$_2$O$_7$	 & 	mp-3932	 & 	0.009	 & 	R140775	 \\
\hline
Hurlbutite	 & 	CaBe$_2$(PO$_4$)$_2$	 & 	mp-6772	 & 	0	 & 	R090048	 \\
\hline
Rynersonite	 & 	CaTa$_2$O$_6$	 & 	mp-18229	 & 	0	 & 	R080064	 \\
\hline
Arsenopyrite	 & 	FeAsS	 & 	mp-561511	 & 	0	 & 	R050071	 \\
\hline
Gudmundite	 & 	FeSbS	 & 	mp-27904	 & 	0	 & 	R060741	 \\
\hline
Cinnabar	 & 	HgS	 & 	mp-634	 & 	0.004	 & 	R070532	 \\
\hline
Cinnabar	 & 	HgS	 & 	mp-9252	 & 	0.004	 & 	R070532	 \\
\hline
Kalsilite	 & 	KAlSiO$_4$	 & 	mp-8355	 & 	0.002	 & 	R060801	 \\
\hline
Kalsilite	 & 	KAlSiO$_4$	 & 	mp-9480	 & 	0.002	 & 	R060030	 \\
\hline
Avogadrite	 & 	KBF$_4$	 & 	mp-4929	 & 	0	 & 	R110062	 \\
\hline
Kotoite	 & 	Mg$_3$(BO$_3$)$_2$	 & 	mp-5005	 & 	0	 & 	R060940	 \\
\hline
Natrosilite	 & 	Na$_2$Si$_2$O$_5$	 & 	mp-3193	 & 	0	 & 	R060855	 \\
\hline
Molybdomenite	 & 	PbSeO$_3$	 & 	mp-20716	 & 	0	 & 	R140388	 \\
\hline
Valentinite	 & 	Sb$_2$O$_3$	 & 	mp-2136	 & 	0	 & 	R120096	 \\
\hline
Stibnite	 & 	Sb$_2$S$_3$	 & 	mp-2809	 & 	0	 & 	R120137	 \\
\hline
Moissanite	 & 	SiC	 & 	mp-7631	 & 	0	 & 	R150016	 \\
\hline
Tellurite	 & 	TeO$_2$	 & 	mp-2125	 & 	0	 & 	R070606	 \\
\hline
Rutile	 & 	TiO$_2$	 & 	mp-2657	 & 	0.037	 & 	R060745, R120008	 \\
\hline
Brookite	 & 	TiO$_2$	 & 	mp-1840	 & 	0.02	 & 	R050363, R050591, R130225	 \\
\hline
Lorándite	 & 	TlAsS$_2$	 & 	mp-4988	 & 	0	 & 	R110055	 \\
\hline
Tungstenite	 & 	WS$_2$	 & 	mp-224	 & 	0	 & 	R070616	 \\
\hline
Waimirite-(Y)	 & 	YF$_3$	 & 	mp-2416	 & 	0	 & 	R130714	 \\
\hline
Reinerite	 & 	Zn$_3$(AsO$_3$)$_2$	 & 	mp-27580	 & 	0	 & 	R080132	 \\
\hline
Baddeleyite	 & 	ZrO$_2$	 & 	mp-2858	 & 	0	 & 	R100171	 \\
\hline
\caption{Common structures in our database based on the same chemical formula and the mineral name in RRUFF compared to Materials Project tags. The bold RRUFF IDs refer to structures that have also similar lattice parameters.} \label{table:2} \\
\end{longtable}

\bibliography{ref}